\documentclass[twocolumn,aps,showpacs,epsf]{revtex4}
\usepackage{graphicx}
\usepackage{dcolumn}
\usepackage{bm}
\def\phd#1#2#3{{ Physica } {\bf #1D}, #2 (#3)}
\def\prl#1#2#3{{ Phys. Rev. Lett.} {\bf #1}, #2 (#3)}
\def\pla#1#2#3{{ Phys. Lett. } {\bf #1A}, #2 (#3)}
\def\pra#1#2#3{{ Phys. Rev. A} {\bf #1}, #2 (#3)}
\def\pre#1#2#3{{ Phys. Rev. E} {\bf #1}, #2 (#3)}

\def\am{amplitude death~}
\def\le{Lyapunov exponent~}
\def\ep{\epsilon~}
\def\ie{i.e.~}
\def\eg{e. g.~}
\def\etl{$et ~al.$~}
\def\etc{ etc.~}

\def\ep{\epsilon}
\def\eqn{\end{equation}\noindent}
\def\eqnr{\end{eqnarray}\noindent}
\def\beqr{\begin{eqnarray}}
 
\begin{document}

\title{Amplitude death in coupled chaotic oscillators}
\author{Awadhesh Prasad}
\email{awadhesh@physics.du.ac.in}
\affiliation{Department of Physics and Astrophysics\\
University of Delhi, Delhi 110 007, India}
\begin{abstract}
Amplitude death can  occur in chaotic dynamical systems with time-delay coupling,
similar to the case of coupled limit cycles. The coupling leads to 
stabilization of  fixed points of the subsystems. This phenomenon is 
quite general, and occurs for identical as well as nonidentical 
coupled chaotic systems.  Using the Lorenz and R\"ossler chaotic 
oscillators to construct representative systems, various possible 
transitions from chaotic dynamics to fixed points are discussed. 
\end{abstract}


\maketitle
\section{Introduction}

{\it Amplitude death}  occurs when coupled oscillators drive each other to fixed points and stop oscillating \cite{eli,am,stro}. Initial studies of this phenomenon were on systems wherein the interaction between the subsystems were  assumed instantaneous. In the absence of coupling the dynamics in the 
individual subsystems were on limit cycles of different periods, near Hopf bifurcations. New stable fixed points are created by the coupling, and these become the attractors of  the dynamics.   

There has been considerable recent interest \cite{general} on the effects of coupling in a variety of  nonlinear systems, not only in the physical sciences, but also in biological and social systems 
in which  in addition to the possibility of  chaotic dynamics, complex phenomena such as synchronization, spatiotemporal intermittency, hysteresis \etc  have been investigated. Theoretical as well as experimental studies of coupled systems have addressed a number of issues which include phase-shifting, phase locking \cite{am} as well as amplitude death.  

It is often necessary to take account of inherent time delays in the coupling, and  Reddy \etl \cite{reddy} investigated the collective dynamical behavior of limit--cycle oscillators interacting diffusively
through time delay at a Hopf bifurcation (see also Sec. \ref{sec:reddy}).  They observed amplitude death of oscillations, regardless of the frequencies of individual oscillators, both theoretically and experimentally. In other work,  amplitude death phenomenon has been studied  theoretically in distributed time-delay \cite{atay} and in a ring of delayed-coupled \cite{ring} oscillators and experimentally in  coupled electronic circuit \cite{reddy2} and coupled thermo-optical oscillators \cite{thermo}. Phenomena similar to amplitude death are also observed in time-delayed self-feedback in limit-cycle oscillators \cite{reddy3} and chaotic oscillators \cite{chaos}.

That this phenomenon can occur for coupled {\it chaotic} oscillators as  well is the main result of the present work. By introducing time--delay  in the coupling of  chaotic subsystems,  irrespective of their type,  one can achieve amplitude death. This transition can be direct, namely from chaotic motion to fixed points, or indirect, when first a pair of limit cycles (rather than a pair of fixed points) are stabilized through the coupling. The motion then asymptotically becomes periodic, and the subsystems oscillate at a common frequency. As a function of the time--delay, however, this limit cycle motion can go from being in--phase to being out--of--phase. This transition is signaled by  a dramatic change in two indicators: the phase difference between oscillators, and their common frequency. Further variation of the time-delay eventually leads to amplitude death via stabilization of new fixed points. 

In the following section we show that identical coupled chaotic systems with matching or mismatched parameters,  and indeed even completely different chaotic systems, all show amplitude death. The in--phase to out--of--phase transition in coupled limit cycle oscillators can be treated analytically. We estimate the common frequency  for a particular case in Section III, and although similar  analysis cannot be carried out for chaotic oscillators, this provides some insight into how the  phenomenon arises more generally. The paper concludes with a discussion and  summary in Section IV. 

\section{Amplitude death}

The basic phenomenology of amplitude death in limit--cycle systems has been  described in some detail \cite{am,reddy}. We consider the general case  of chaotic coupled systems,
\begin{eqnarray}
\dot{\bf x} &=& {\bf f_x}{\bf (x)} + {\bf g_x(y(}t-\tau){\bf,x(}t{\bf))}
\nonumber\\
\dot{\bf y} &=& {\bf f_y}{\bf (y)}+ {\bf h_y(x(}t-\tau){\bf,y(}t{\bf))}
\end{eqnarray}
where {\bf x} and {\bf y} denote the variables of the two subsystems.  The dynamical equations are specified by ${\bf f_x}$ and ${\bf f_y}$  respectively, and ${\bf g_x}$ and ${\bf h_y}$ specify the couplings,  $\tau$ being the time--delay.

Several scenarios are possible. In the simplest case, the subsystems  specified by the variables ${\bf x}$ and ${\bf y}$ can be identical.  This is dealt with in subsection A below. By altering the parameters, 
the two subsystems can be made nonidentical, and this case is treated in subsection B. Finally, the two chaotic systems can be completely different, and this case is presented in subsection C. In all cases 
we demonstrate that amplitude death is possible, though the mechanism in each case is somewhat different.  For simplicity, the results are presented through applications to specific R\"ossler  or Lorenz dynamical systems which have been extensively studied in the context of chaotic dynamics \cite{attlee}. The present results are mainly numerical since analytic results for such nonlinear systems are difficult to obtain. 

In each of the cases, we address the possibility of amplitude suppression, namely the stabilization of limit cycles. For this phenomenon, some analysis is possible, and is presented in Sec. \ref{sec:reddy}.

\subsection{Identical Chaotic Subsystems}

The R\"ossler system \cite{rossler} is a simple mathematical model  of chemical kinetics that incorporates  reaction-diffusion, and  has been extensively studied in the past two decades as one of  the simplest chaotic flows.  Consider  two identical R\"ossler oscillators denoted by variables ${\bf x}$ and ${\bf y}$ 
\begin{eqnarray}
\nonumber
 d x_1(t)\over dt &=& - x_2 -x_3\\
\nonumber
 d x_2(t) \over dt &=& x_1 +a x_2 + \nonumber
 \epsilon[y_2(t-\tau)-x_2(t)]\\ \nonumber
d x_3(t)\over dt &=& b+ x_3 (x_1 - c_1)\\ \nonumber
 d y_1(t)\over dt &=& - y_2 -y_3\\ \nonumber
 d y_2(t) \over dt &=& y_1 +a y_2 
 +\epsilon[x_2(t-\tau)-y_2(t)]\\
d y_3(t)\over dt &=& b+ y_3 (y_1 - c_2)  
\label{eq:ross}
\end{eqnarray}\noindent
which are  symmetrically coupled through $x_2$ and $y_2$. We take numerical values of the  parameters as $a=b=0.1$ and  $c_1=c_2=c=14$  with the coupling parameters $\epsilon$ and the delay
$\tau$ being treated as variable. 

\begin{figure}
\scalebox{0.55}{\includegraphics{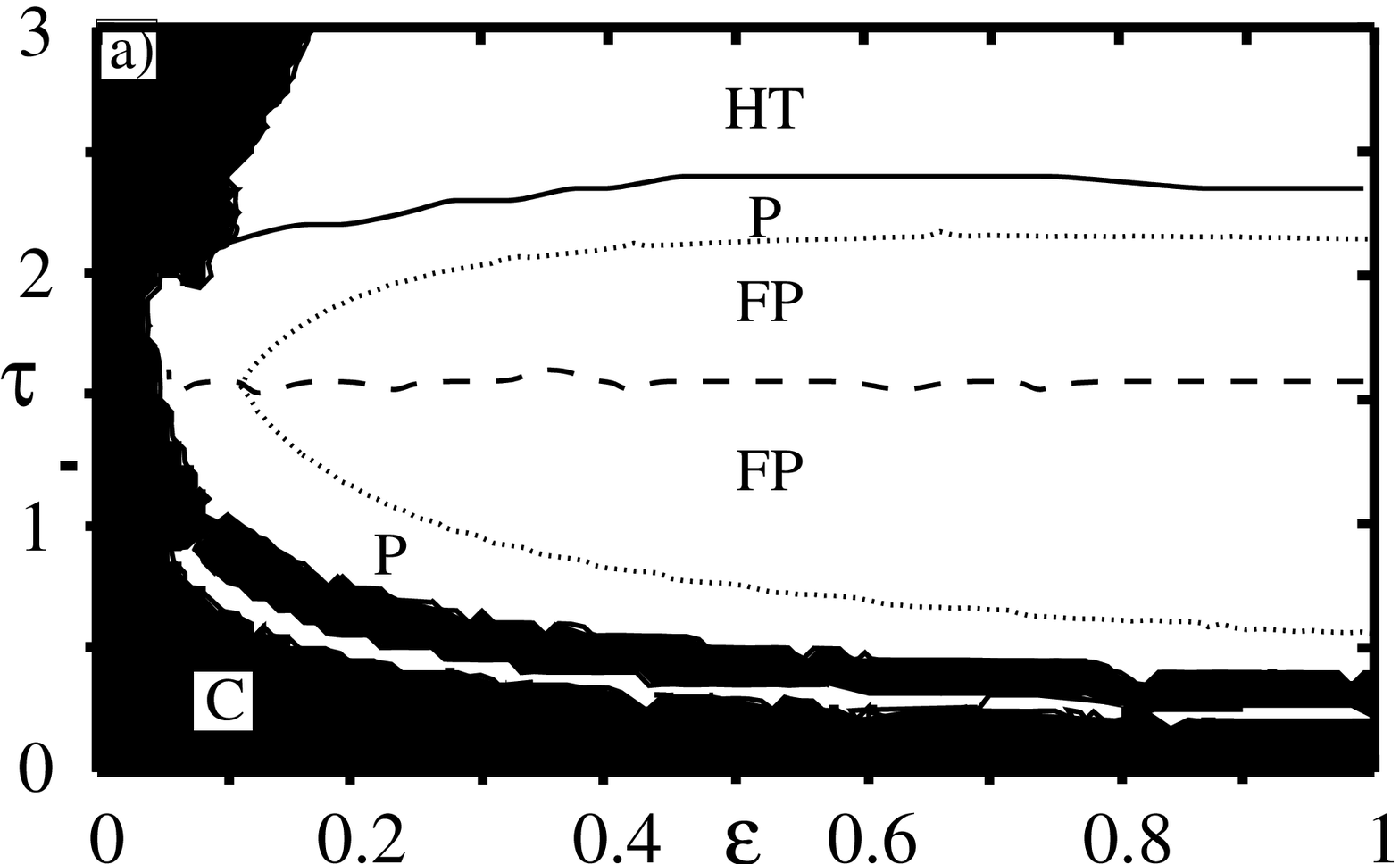}}
\vskip1.5cm
\scalebox{0.55}{\includegraphics{figure1b}}
\caption{(Color online)(a) Schematic phase diagram of identical
R\"ossler oscillators, Eq.~(\ref{eq:ross}), in the
$\ep-\tau$ plane for $c_1=c_2=18$.  In the dark region the motion
is  chaotic ($C$) while in the white region there are
regular states of different kind \eg fixed point  ($FP$),
periodic cycles (P) and hypertorus (HT). Numerical details
are given in \cite{num}. The dotted line indicates the locus where
largest \le ($\lambda_1$) becomes negative. 
(b) The spectrum of  Lyapunov exponents (black, green and blue
correspond to $\lambda_1$, $\lambda_2$, and $\lambda_3$ respectively)
as a function of time delay $\tau$ at fixed coupling strength $\ep=0.5$.
The vertical  arrow in (a) shows the parameter value, $\tau_c$,
when the motion goes from being in phase to out of phase.
The largest Lyapunov exponent of the fixed point is shown in red. 
In the \am range all the Lyapunov exponents are negative and 
$\lambda_1=\lambda_2= \lambda_{FP}$. The transition is from
limit cycle to a fixed point. 
(c) The phase difference between oscillators with time delay $\tau$,
which before and after the transition are equal to $0$ and $\pi$.  
(d) Numerically calculated common frequency, $\Omega$, 
as a function of the time delay $\tau$.}
\label{fig:ident}
\end{figure}

In absence of coupling,  $\ep=0$, the subsystems, which are chaotic  \cite{rossler}, evolve independently. For finite coupling strength  various states of motion are observed. These are 
indicated  in the schematic Fig.~\ref{fig:ident}(a) within a representative range of coupling parameters $\ep$ and $\tau$.  The  region marked $C$ which is shaded black, corresponds to   chaotic states  while the white region shows the regions  of regular behavior, namely  periodic (P), fixed point (FP) 
and hypertorus (HT)  \cite{ht} dynamics.  Numerical details are given in \cite{num,farmer}. Finer analysis at a fixed value of coupling strength, $\ep=0.5$ is  shown in Fig.~\ref{fig:ident}(b). The three largest 
Lyapunov exponents are shown as function of the time delay $\tau$.  The dotted line in Fig.~\ref{fig:ident}(a) indicates the  transition  where the largest \le ($\lambda_1$) becomes negative  [see Fig.~\ref{fig:ident}(b)].  The vertical arrow in Fig.~\ref{fig:ident}(b) denotes the  parameter at which the slope of  $\lambda_1$ changes sign. In Fig.~\ref{fig:ident}(a), the locus of  this point  is plotted as a dashed line.  The solid line in  Fig.~\ref{fig:ident}(a) shows the transition from periodic solution   to the hypertorus where  $\lambda_1 \approx \lambda_2 \approx 0$.

\begin{figure}
\scalebox{0.70}{\includegraphics{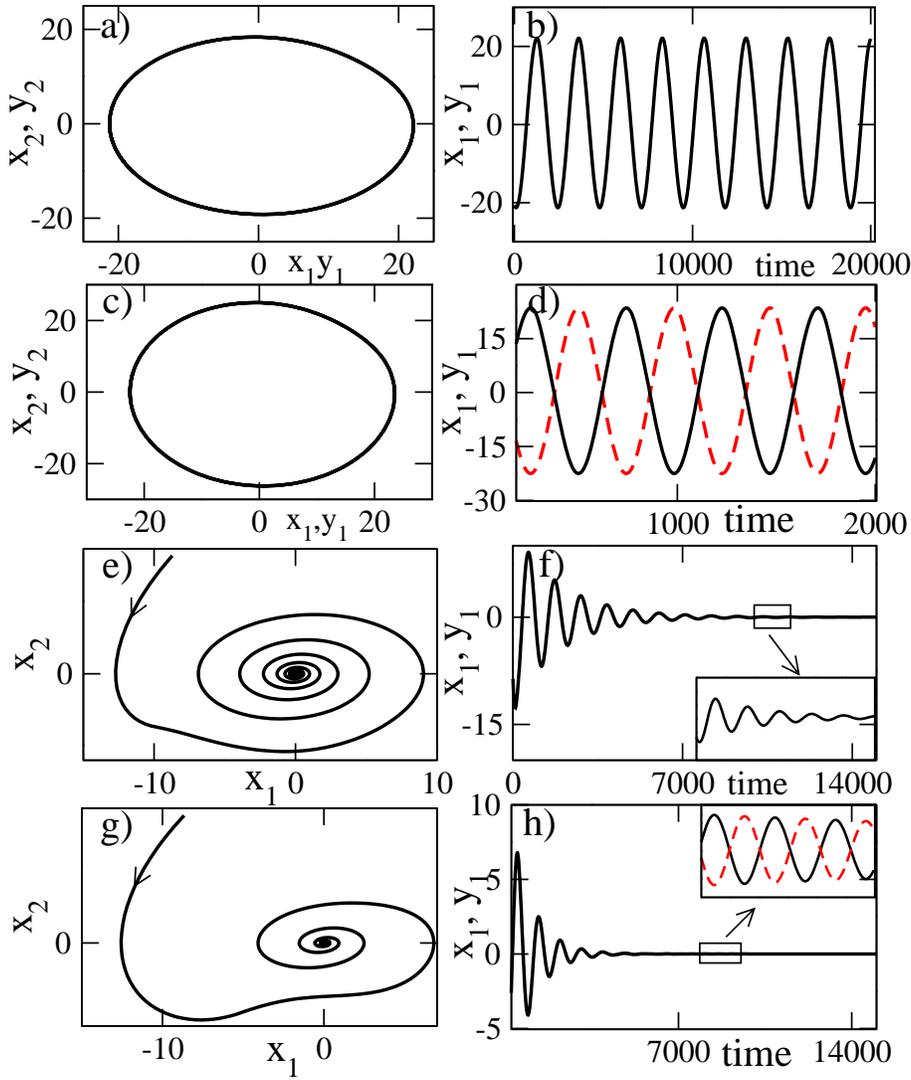}}
\caption{Trajectories of oscillator 1 (solid line) and 2 (dashed line)
in phase space (left panel) and with time 
(right panel) (a) and (b) at 
$\tau=0.6$, (c) and (d) at  $\tau=2.25$, (d) and (e) at
 $\tau=1.5$, and, (g) and (h) at $\tau=2$ respectively.}
\label{fig:ident_traj}
\end{figure}

In the region marked  P in Fig.~\ref{fig:ident}(b), the dynamics of the two subsystems settles onto limit cycles. Trajectories \cite{traj} of the individual  oscillators in the $x_1-x_2$ plane with $\tau =0.6$ are shown in  Figs.~\ref{fig:ident_traj}(a) and (b) respectively.  Here the trajectories of both the oscillators overlap since they are in complete synchronization. Further, these are also in phase since  the phase difference is zero. There is also another periodic  region between the FP and HT motions  (between dotted and solid lines in  Fig.~\ref{fig:ident}(a)) where trajectories of  individual oscillators  at $\tau=2.25$ are shown in  Figs.~\ref{fig:ident_traj}(c) and (d). In  the latter case  both the oscillators show the same periodic  motion (frequency locked states) (see Fig.~\ref{fig:ident_traj}(c))  but they are out of phase [Fig.~\ref{fig:ident_traj}(d)], with  a constant phase difference of $\pi$. 

Both types of periodic motions (either in or out of phase) can arise from the following reasons  \cite{pyra}. Either the coupling  stabilizes  one of the (infinitely  many) unstable  periodic orbits (UPO) \cite{upo}  which are  embedded in the chaotic set, or creates new periodic solutions with  time period ($T$) different from the time delay, \ie $\tau\ne T$  (the perturbation term  does not vanish here).   In the former case only those UPOs will be stabilized whose time periods   are equal to the time delay, namely $\tau=T$. A trajectory will settle  on an UPO and  remain there since the feedback function becomes zero:
the coupling term vanishes when $y_2(t-\tau)$ becomes equal to $x_2(t)$.   This implies that the  perturbation does not change the original solution.  In the latter situation, the effect of time delay is shown in  Figs.~\ref{fig:ident_traj}(a-d).

As $\tau$ is increased further (between the dotted lines  in  Fig.~\ref{fig:ident}(a), where the largest Lyapunov exponent,  $\lambda_1$ is negative as can be seen in  Fig.~\ref{fig:ident}(b)) fixed point (FP) solutions are obtained. Examples of such solutions,  before and after the marked arrow in Fig.~\ref{fig:ident}(b)  are  shown in  Figs.~\ref{fig:ident_traj}(e,f) and (g,h) respectively for $\tau =1.5$ and $\tau=2$.  In  both cases transient  trajectories start spiraling into one of the fixed points $x_{1\star}=y_{1\star}=ax_{3\star},  x_{2\star}=y_{2\star}=-x_{3\star}, x_{3\star}=y_{3\star},  x_{3\star}=[c-\sqrt(c^2-4ab)]/2a$.  In  these regions, since the subsystems are  identical, the coupling term vanishes and the  fixed points remain the same as  that of the uncoupled systems.  The largest Lyapunov exponent for the  fixed point solution  ($\lambda_{FP}$), the position of which  does not change with  time delay at a given value of the  coupling strength,  is also shown in Fig.~\ref{fig:ident}(b) (red line with circles). 
Since this is negative in the FP region (where $\lambda_1=\lambda_2=\lambda_{FP}$),  the fixed point
is stable. Any attempt by the subsystems to move from this state increases the coupling term and brings the entire system back to  the fixed point. Thus by using time delay coupling, amplitude death 
phenomena can be observed even in chaotic systems in a manner which  is very similar to that in limit--cycles \cite{am,reddy}. 

In the transition  from chaos to amplitude death that occurs  via a limit cycle, the mechanism is the same as that of the coupled limit cycle case \cite{am,reddy} (see also  Sec. \ref{sec:reddy}) \ie the transition from limit cycle to FP takes  place because a pair of complex eigenvalues cross the imaginary 
axis from right to left and the Lyapunov exponent $\lambda_{FP}$ becomes negative.  Simultaneously the stability  of the fixed point is lost as a pair of eigenvalues cross  the axis from  left to right.  Thus, in this case amplitude death is initiated at a Hopf bifurcation.

There are however some differences in the manner in which the fixed  points are reached. The largest Lyapunov exponent has a change in slope at the point $\tau_c$  marked by the arrow, and differences can  be seen in the phase relationship of the two oscillators:  for $\tau  <   \tau_c$ [inset of Fig.~\ref{fig:ident_traj}(f)] the two are in--phase, while for $\tau  >   \tau_c$ [inset of  Fig.~\ref{fig:ident_traj}(h)], they are out of phase. Shown in Fig.~\ref{fig:ident}(c) is the phase difference ($\Delta \theta$)
between oscillators which is defined as $\Delta \theta = \langle|\theta_1(t)-\theta_2(t)|\rangle$ where $\langle\cdot\rangle$ denotes the average over  time while $\theta_1(t)\sim \tan^{-1}[(x_2(t)/x_1(t)]$ and 
$\theta_2(t)\sim \tan ^{-1}[y_2(t)/y_1(t)]$ \cite{phasediff,trans}. This clearly indicates that below $\tau_c$ the phase difference is  zero while after it is $\pi$.  We also observe (see Figs.~\ref{fig:ident_traj}(b,d,f,h)) that the two oscillators are frequency--locked. Their common frequency, $\Omega$,  (measured from the peak-to-peak separation \cite{trans}) is shown in Fig.~\ref{fig:ident}(d) where a step--like change is observed at $\tau_c$.  This behaviour is analyzed in a simpler system in the following Section.

The change in the phase difference and common frequency occurs at parameter values corresponding to maximal stability, namely when the Lyapunov exponent is at a local minimum. While details of
the precise mechanism in chaotic systems needs further analysis, it can be easily verified that this behavior is quite general, and  a similar transition occurs in a variety of model systems.

\begin{figure}
\scalebox{0.50}{\includegraphics{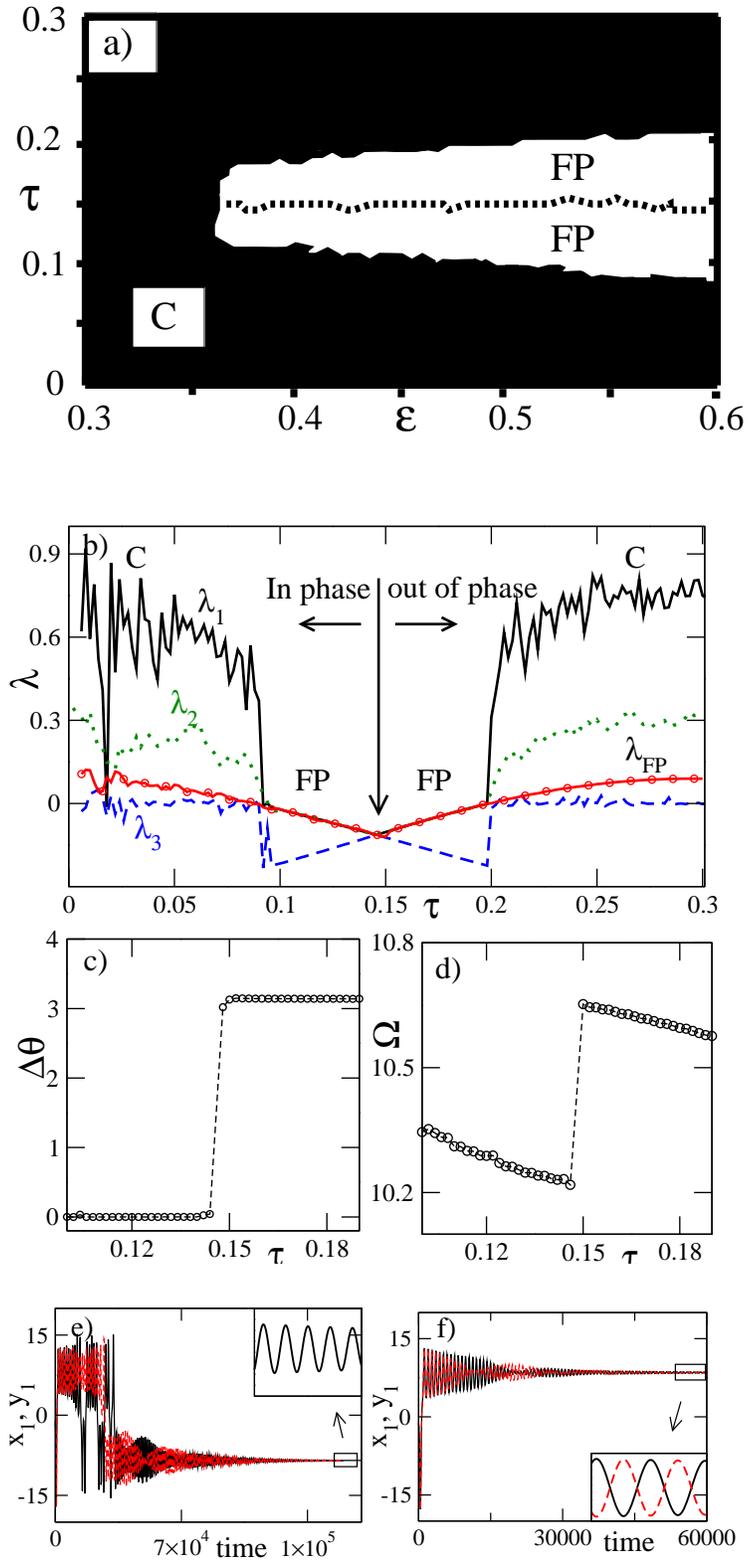}}
\vskip1.2cm
\scalebox{0.55}{\includegraphics{figure3b}}
\vskip.4cm
\scalebox{0.51}{\includegraphics{figure3c}}
\caption{(Color online) (a) Schematic phase diagram for the
Lorenz system Eq.~(\ref{eq:lorenz}) in $\ep-\tau$ plane.
(b) Spectrum of Lyapunov exponents (black, green and blue
corresponding to $\lambda_1$, $\lambda_2$, and $\lambda_3$ respectively)
as a function of the time delay, $\tau$, at fixed coupling strength $\ep=0.5$.
The other details remains the same as that in the Fig.~\ref{fig:ident}.
In the \am range all the Lyapunov exponents are negative and 
$\lambda_1=\lambda_2= \lambda_{FP}$; the transition is directly from 
chaos to a fixed point.
(c) The phase difference between oscillators with time delay $\tau$.
The phase difference before and after the transition are equal to $0$
 and $\pi$.  (d) Numerically calculated common frequency, $\Omega$.
Trajectories of oscillators 1 (solid line) and 2 (dashed line)
  with time at  (e) 
$\tau=0.12$ and  (f)  $\tau=0.17$.  The inset figures
are the  enlarged view of the corresponding marked boxes.}
\label{fig:loren}
\end{figure}

Consider coupled identical chaotic Lorenz oscillators \cite{lorenz},
\begin{eqnarray}
\nonumber
 d x_1(t)\over dt &=& -\sigma (x_1-x_2)\\
\nonumber
 d x_2(t) \over dt &=& -x_1 x_3-x_2+ r_1 x_1
 +\epsilon [y_2(t-\tau)-x_2(t)]\\
\nonumber
d x_3(t)\over dt &=&x_1 x_2 -dx_3\\
\nonumber
 d y_1(t)\over dt &=& -\sigma (y_1-y_2)\\
\nonumber
 d y_2(t) \over dt &=& -y_1 y_3-y_2+ r_2 y_1
 +\epsilon [x_2(t-\tau)-y_2(t)]\\
d y_3(t)\over dt &=&y_1 y_2 -dy_3
 \label{eq:lorenz}
\end{eqnarray}
with $\sigma=10, r_1=r_2=28, d=8/3$. The phase diagram as a function of  the  coupling parameters $\tau$ and $ \ep$ is shown in  Fig.~\ref{fig:loren}(a), while the  Lyapunov exponents with time  delay $\tau$ at fixed  $\ep=0.5$ are shown in Fig.~\ref{fig:loren}(b). The step transition in phase difference  and common frequency are  shown in Fig.~\ref{fig:loren}(c) and (d) respectively.  The  behaviour  is very similar to  the R\"ossler case, Fig.~\ref{fig:ident}. Transient  trajectories that go to the fixed point across the marked  vertical arrow  at $\tau=0.12$ and $\tau=0.17$ are shown in   Figs.~\ref{fig:loren}(e) (in--phase) and  Figs.~\ref{fig:loren}(f) (out of phase) respectively. The fixed point at ($\pm\sqrt{d(r-1)},\pm\sqrt{d(r-1)},r-1$) is stable in the region marked FP, and has a negative Lyapunov exponent, $\lambda_{FP}$ \cite{lor}. In this region $\lambda_1,=\lambda_2=\lambda_{FP}$. Thus in this regime amplitude death  also occurs.

However there is difference in manner of transition from that of  Fig. \ref{fig:ident}. In the case of coupled limit-cycle \cite{reddy} or R\"ossler system, Eq.~(\ref{eq:ross}), transition is from periodic to fixed point while in this case it  happens directly from chaotic dynamics to a fixed point. This implies that  although neither UPOs are stabilized, nor new periodic solutions are created via the time-delay interaction,
the fixed point is stabilized directly and this leads to an abrupt change in the largest \le of the system, $\lambda_1$.  Such a transition, which has not been discussed earlier, suggests that a new mechanism may be operative in amplitude death.

\subsection{Nonidentical Chaotic Subsystems}

\begin{figure}
\scalebox{0.5}{\includegraphics{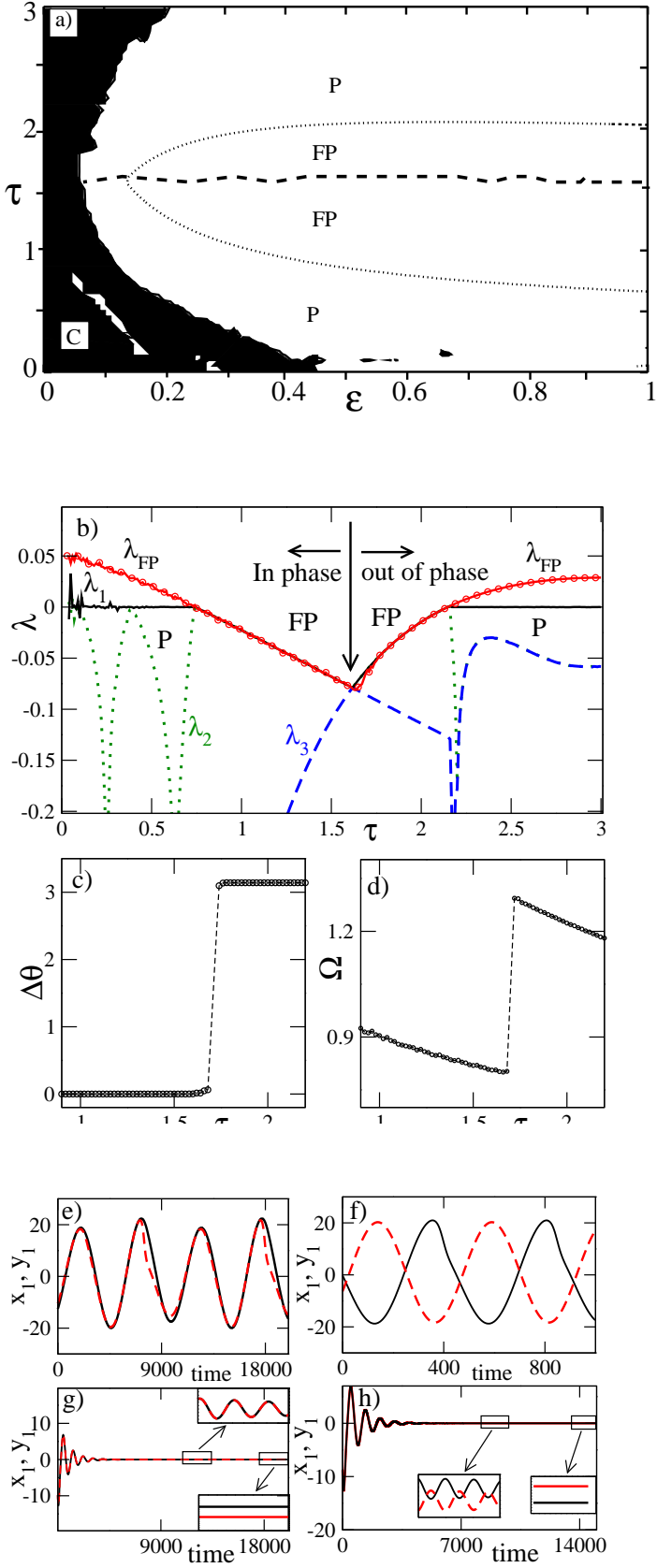}}
\vskip1.3cm
\scalebox{0.5}{\includegraphics{figure4b}}
\vskip.8cm
\scalebox{0.5}{\includegraphics{figure4c}}
\caption{ (Color online) (a) Schematic phase diagram of
 Eq.~(\ref{eq:ross}) in $\ep-\tau$ plane.
Parameters values are $c_1=14$ and $c_2=18$.
(b) Spectrum of Lyapunov exponent  (black, green and blue
correspond to $\lambda_1$, $\lambda_2$, and $\lambda_3$ respectively) 
as a function of the time delay, $\tau$, at fixed coupling strength $\ep=0.5$.
Other details remains the same as that of Fig.~\ref{fig:ident}.
In  \am range  all Lyapunov exponents are negative and 
$\lambda_1=\lambda_2= \lambda_{FP}$. 
This is also a chaos to FP transition as in Fig.~\ref{fig:ident}.
(c) The phase difference between oscillators with time delay
 $\tau$.
The phase difference before and after the transition are near to
$0$ and $\pi$ respectively.  (d) Numerically calculated common frequency, 
$\Omega$.
Trajectories of oscillator 1 (solid line) and 2 (dashed line)
for the nonidentical case  with
time  (e)  at $\tau=0.25$,
(f) at  $\tau=2.5$, (g) at  $\tau=1$,
 and, (h) at $\tau=2$. The inset figures
are  the enlarged views of the corresponding  marked boxes.}
\label{fig:diff}
\end{figure}

We consider the effect of a difference in parameters between the  two subsystems. This is  particularly important with respect to  experimental realization of such phenomena since in practice it is 
impossible to ensure that all parameters of two different systems  are exactly equal. Introduce a mismatch in the parameters of the R\"ossler system,  Eq.~(\ref{eq:ross}). For  $c_1=14$ and $c_2=18$, various possible  states are shown in Fig.~\ref{fig:diff}(a). Other parameters are  the same as in  Fig. \ref{fig:ident}. A major difference, compared with Fig.~\ref{fig:ident} is the absence of hypertorus dynamics. The spectrum of Lyapunov exponents  at $\ep=0.5$ are shown in  Fig.~\ref{fig:diff}(b); in the fixed point region all the Lyapunov exponents are negative. 
The periodic states at $\tau=0.25$ and $\tau=2.5$ are shown in 
 Figs.~\ref{fig:diff}(e) and (f) where the motions are in and out of phase respectively (these are in the frequency locked regime where time periods are same for both the oscillators).

Shown in Figs.~\ref{fig:diff} (g) and (h) are transient  trajectories at $\tau=1$ and $\tau=2$ respectively. The insets are enlarged views of the corresponding transient and fixed point motions.  Here coupling introduces a {\it new} set of fixed points  and  the system  settles in one of these. The fixed points for 
Eq.~(\ref{eq:ross}) simply turn out to be  
\beqr 
x_{1\star}&=&ax_{3\star}-\ep(x_{3\star}-y_{3\star}),\nonumber\\
 x_{2\star}&=& -x_{3\star},\nonumber\\ 
x_{3\star}&=& [(\ep-a)y_{3\star}^2+c_2y_{3\star} -b]/\ep y_{3\star},\nonumber\\
y_{1\star}&=&ay_{3\star}- \ep(y_{3\star}-x_{3\star}),\nonumber\\
 y_{2\star}&=&-y_{3\star},
 \eqnr
 where   $y_{3\star}$ is the root of the polynomial
$$ b+y_{3\star}[ay_{3\star}-\ep(y_{3\star}-x_{3\star})-c_2]=0.$$ The largest Lyapunov exponent
 $\lambda_{FP}$ (red line with circles), of the fixed point  $x_{1\star}= 0.00135.., x_{2\star}=-0.0056..,
x_{3\star}=0.0056.., y_{1\star}=-7.914..\times 10 ^{-5}, y_{2\star}=-0.00714..,$ and $y_{3\star}=0.00714..$ is shown in Fig.~\ref{fig:diff}(b).  Here the transition is from chaos to a fixed point occurs indirectly, namely via a limit cycle at a Hopf bifurcation as in the case of identical coupled R\"ossler oscillators, Fig.~\ref{fig:ident}.

Similar step  transitions in the phase difference and common 
frequency  as in Figs. \ref{fig:ident}(c,d)  and \ref{fig:loren}
 (c,d) are also observed here in Figs. \ref{fig:diff}(c,d). However the phase difference are  only $\approx 0$ and $\approx \pi$ before and after the transition (marked by the arrow).

Amplitude death phenomena thus also occurs in nonidentical chaotic subsystems, though there can be subtle differences in the nature of the final state which can have either in--phase \cite{reddy,am} 
or out of phase \cite{phase}  similar to the identical subsystem case.  Similar results have been obtained \cite{new} for the case of mismatched coupled Lorenz systems.

\subsection{Mixed Chaotic Subsystems}

Since amplitude death occurs for interacting oscillators with very different time--periods, it is natural to consider the situation when the two subsystems are dynamically distinct. This is of even greater relevance experimentally, since a variety of different oscillator systems can be mutually coupled
in a given natural situation.

\begin{figure}
\scalebox{0.45}{\includegraphics{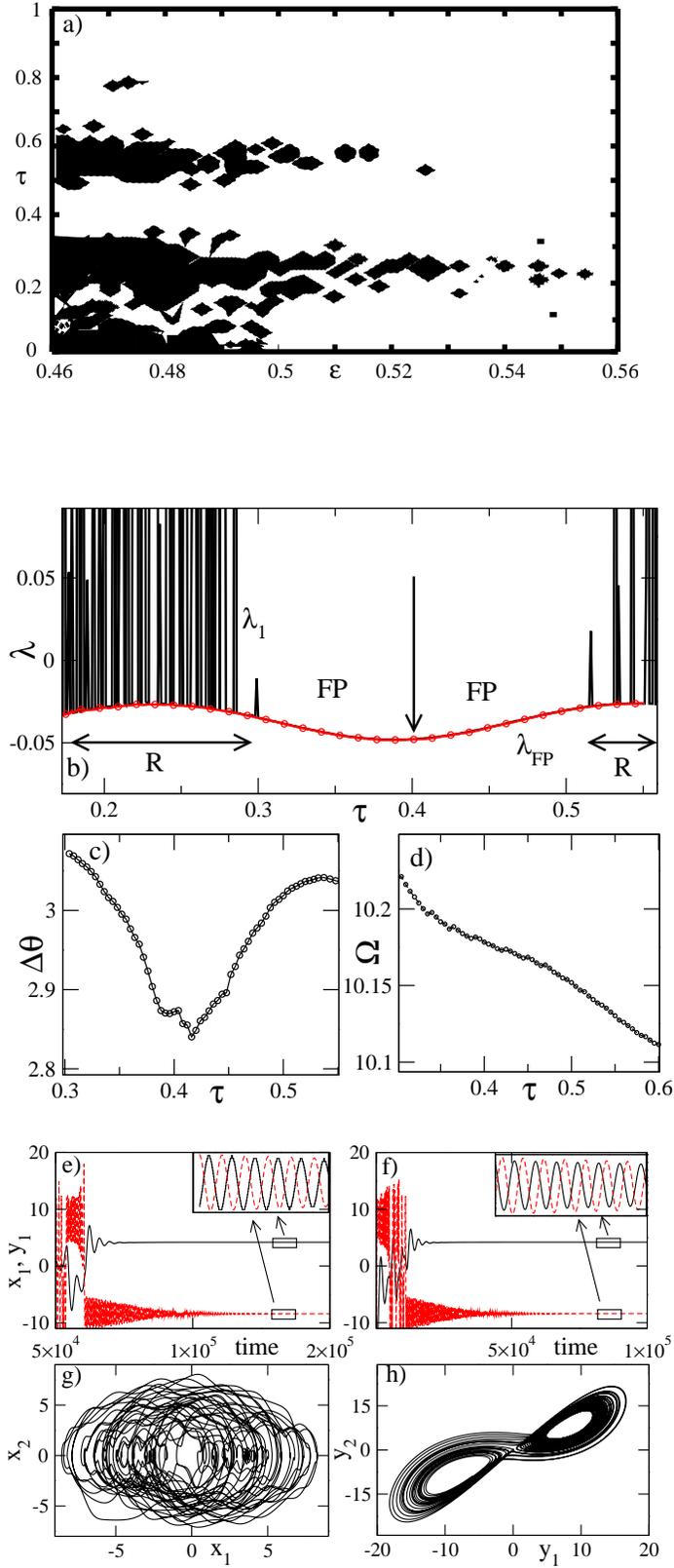}}
\vskip1.7cm
\scalebox{0.5}{\includegraphics{figure5b}}
\vskip.5cm
\scalebox{0.450}{\includegraphics{figure5c}}
\caption{(Color online) (a) Schematic phase diagram for the mixed 
R\"ossler and Lorenz system, Eq.~(\ref{eq:mix}),
 in $\ep-\tau$ plane.
  (b) The largest  Lyapunov exponent (in black) as a function of 
time delay, $\tau$, at fixed coupling strength $\ep=0.5$.
$R$ denotes regions of wild fluctuations in the Lyapunov
exponent where chaotic and stable fixed point solutions coexist.
The red line with circles represents the largest Lyapunov exponent of a fixed point 
($x_{1\star}=4.200308.., x_{2\star}=-0.007246.., x_{3\star}=-x_{2\star},
y_{1\star}=-8.406415.., y_{2\star}=y_{1\star},$ and $y_{3\star}=
26.500431..$). In contrast to previous examples, 
there is no discontinuous change in slope in the largest Lyapunov exponent.
(c)The phase difference between oscillators with time delay
 $\tau$.
 (d) Numerically calculated common frequency, $\Omega$.
Trajectories of R\"ossler (solid line) and
Lorenz (dashed line) oscillators for
(e) $\tau=0.35$ and (f) $\tau=0.45$.
Chaotic trajectories of (g) R\"ossler and
(h)  Lorenz oscillators  at $\tau=0.25$.  The  inset figures,  which
ordinate axes are scaled to see phase relations,
are the  enlarged view of  corresponding marked boxes.}
\label{fig:mix}
\end{figure}

To examine the behavior that obtains when distinct  oscillators are coupled, 
I study coupled  R\"ossler and Lorenz oscillators
\begin{eqnarray}
\nonumber
 d x_1(t)\over dt &=& - x_2 -x_3\\
\nonumber
 d x_2(t) \over dt &=& x_1 +a x_2 + \nonumber
 \epsilon[y_2(t-\tau)-x_2(t)]\\ \nonumber
d x_3(t)\over dt &=& b+ x_3 (x_1 - c_1)\\ \nonumber
 d y_1(t)\over dt &=& -\sigma (y_1-y_2)\\
\nonumber
 d y_2(t) \over dt &=& -y_1 y_3-y_2+ r_2 y_1
 +\epsilon [x_2(t-\tau)-y_2(t)]\\
d y_3(t)\over dt &=&y_1 y_2 -dy_3
 \label{eq:mix}
\end{eqnarray}
where parameters $c_1=18$ and $r_2=28$. Other parameters are the same
as in Eqs.~(\ref{eq:ross}) and (\ref{eq:lorenz}). 
A typical phase diagram  is shown in Fig.~\ref{fig:mix}(a),
and the Lyapunov exponent as a function of delay $\tau$ is shown 
in Fig.~\ref{fig:mix}(b).
Various possible states are indicated: the region of  
fixed point motion, \ie amplitude death  phenomena, is observed in 
the range $\tau \sim0.3- \tau \sim0.52$ (where 
$\lambda_1=\lambda_{FP})$. Trajectories  for 
$\tau=0.35$ and $\tau=0.45$ in \am region are shown in
 \ref{fig:mix}(e) and (f) respectively. The inset figures 
indicate that both the subsystems 
are neither completely in nor completely out of phase 
\ie phase differences are away from zero or near $\pi$ (see Figs. \ref{fig:mix}(c) and (d))  
 due to the completely
distinct natures of the chaotic dynamics of the individual subsystems.
Even though the subsystems are distinct, it has been possible to observe indications of a vestige of the  transition in the phase difference  when \le is at a local minimum \cite{new}.

Regions, denoted by $R$, where wild  fluctuations  (see Fig.~  \ref{fig:mix}(a))  in the Lyapunov exponent occur are indicative of  the presence of coexisting attractors with complicated basins 
\cite{pla2,new}. I have  verified the presence of  riddled \cite{riddle} basins
(results are not presented here) in this region, and identify at least two  coexisting attractors:
the chaotic trajectories (shown separately for R\"ossler and Lorenz 
subsystems in Figs.~\ref{fig:mix}(g) and (h) respectively), and the fixed
point attractor, in  Figs.~\ref{fig:mix}(e) and (f). 
Here the transition is directly from chaos to fixed point  similar to
Lorenz oscillator, Eq.~(\ref{eq:lorenz}) (neither is an UPO 
stablised nor is a new periodic solution created)
but in the present case the FP solution is created via riddling.

\section{Amplitude suppression: Analytical Estimation of the Frequency Jump} \label{sec:reddy}
The phenomenon of frequency jump which can be observed (See Sections IIA, B) in identical R\"ossler and Lorenz chaotic systems can be seen in coupled limit cycles systems as well. Consider the case that has been studied in detail by  Reddy \etl \cite{reddy}, 
\begin{eqnarray}
\label{eq:reddy}
\nonumber
\dot{Z}_1(t) &=& (1+i\omega_1-|Z_1(t)|^2)Z_1(t)+\ep[Z_2(t-\tau)-Z_1(t)]\\
\nonumber
\dot{Z}_2(t) &=& (1+i\omega_2-|Z_2(t)|^2)Z_2(t)+\ep[Z_1(t-\tau)-Z_2(t)]\\
&&\\
\nonumber
\end{eqnarray}
where $Z_j(t), j=1,2$ are the complex amplitudes of $j-$th oscillator
with $|Z_j|=1$ and corresponding angular frequencies $\omega_j$.
$\ep$ is the coupling strength. For simplicity we consider identical 
oscillators with $\omega_1=\omega_2$. 

\begin{figure}
\scalebox{0.70}{\includegraphics{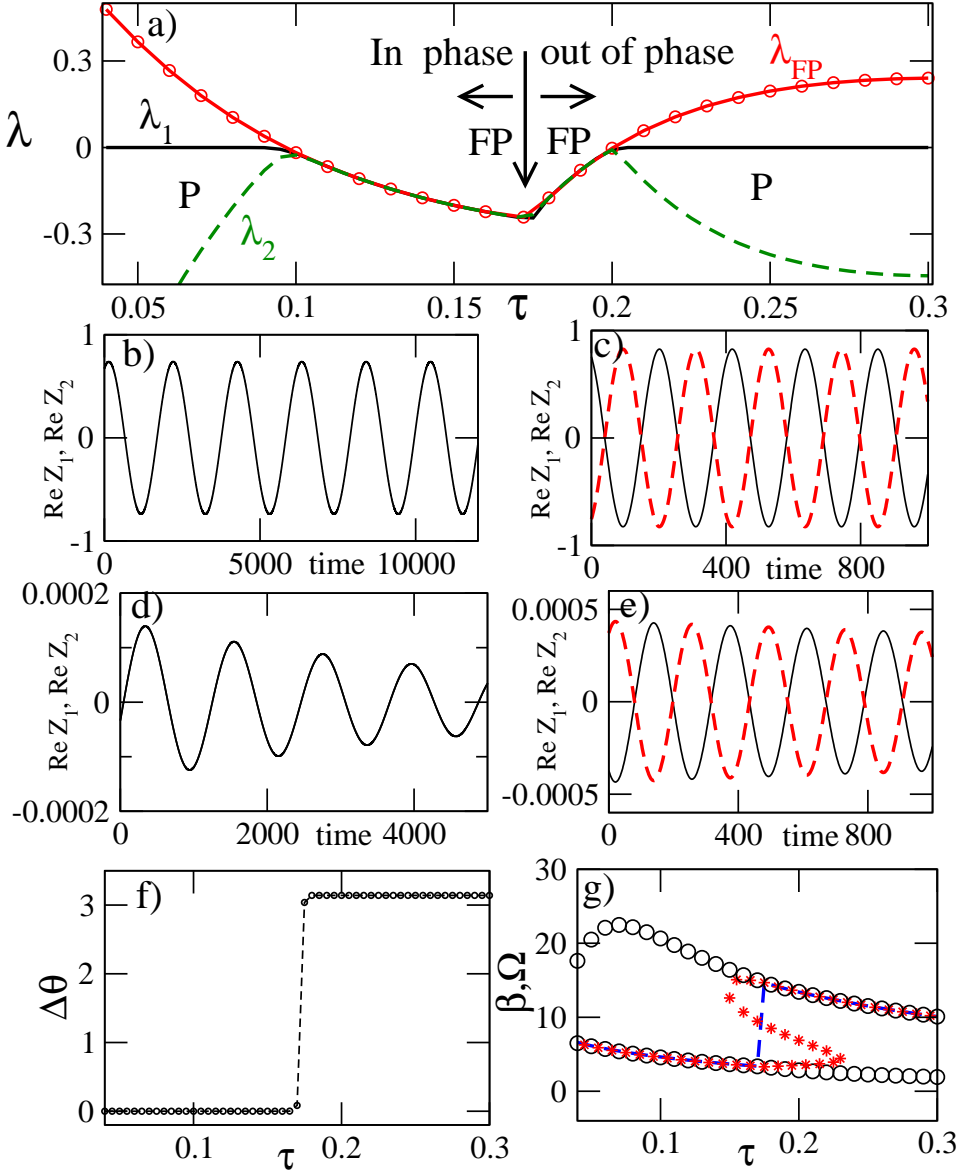}}
\caption{(Color online) (a) Spectrum of Lyapunov exponents
for the coupled limit cycle oscillators (black and green 
correspond to $\lambda_1$ and  $\lambda_2$ respectively) 
with time delay, $\tau$, at fixed coupling strength $\ep=10$.
The red line with circles is the largest Lyapunov exponent for 
the fixed point, $Z_j=0$.  In \am region all Lyapunov exponents 
are negative and $\lambda_1=\lambda_2= \lambda_{FP}$.
Trajectories of the oscillators 1 (solid line) and 2 (dashed line) 
with limit cycles at (b) $\tau=0.05$ (in phase)
and (c) $\tau=.25$ (out of phase).
$Re Z_j$ represents the real part for $Z_j$ of $j-$th oscillator.
Transient trajectories of oscillators 1  (solid line) and
2  (dashed line)  with time at  
(d) $\tau=0.14$ (in phase) and (e) $\tau=0.19$ (out of Phase) 
respectively for fixed point solutions.
(f) The phase difference between oscillators with time delay $\tau$ which
is equal to $0$ and $\pi$ before and after the transition respectively. 
(g) Solution of Eqs.~(\ref{eq:freq1}) (black circles) and
 (\ref{eq:freq2}) (red stars) with parameter $\tau$. Blue 
dashed line indicates the numerically calculated common 
frequency, $\Omega$.}
\label{fig:reddy}
\end{figure}

Specific results are shown for the case of $\omega_1 = 9$, which 
was the case used in Ref. \cite{reddy}.  The origin is a fixed point,
\ie $Z_j=0$. The spectrum of Lyapunov exponents and $\lambda_{FP}$ 
 the exponent (the red line with circles) for the fixed point, for
coupling strength $\ep=10$ is shown in Fig.~\ref{fig:reddy}(a). 
The resulting limit cycles at $\tau=0.05$ are in phase,
and at $\tau=0.25$ are out of phase, and these
are shown in Fig.~\ref{fig:reddy}(b) and (c) respectively. 
The amplitude death region is in between $\tau \sim 
0.1$ and $0.2$ where all Lyapunov exponents are negative, $\lambda_1=
\lambda_2=\lambda_{FP}$.
The transient behaviour of in and out of phase motions
in this amplitude death region are shown in Fig.~\ref{fig:reddy} (d) and 
(e) at $\tau=0.13$ and $0.19$ respectively. The transition form
limit cycle to fixed point motion (say from $\tau=0.05$ to
$\tau=0.14$) is happening at the  Hopf bifurcation.
The Lyapunov exponent of  the fixed point, $\lambda_{FP}$, 
indicates this transition clearly, when it becomes negative.

Similar to Figs. \ref{fig:ident}(c) and \ref{fig:loren}(c),
shown in Fig.~\ref{fig:reddy}(f) is the
phase difference between oscillators of Eq.~(\ref{eq:reddy}).
The  phases of the individual oscillators are defined here
 as $\theta_j=\tan^{-1}[Im(Z_j)/Re(Z_j)]$. The phase difference
changes drastically from $0$ to $\pi$ when  the largest \le~
is minimum (the system is in most stable stat). 

There is also a similar step transition in  frequency
across which in-- and out--of---phase motions are present.
The numerically  calculated frequency, $\Omega$, of Eq.~(\ref{eq:reddy})
with time delay, is shown by the dashed line in Fig.~\ref{fig:reddy}(g).
In order to understand this transition I consider the 
 characteristic eigenvalue equation for Eq.~(\ref{eq:reddy}),
by assuming  the linear perturbation to vary as $e^{\lambda t}$, which
can be written as 
\cite{reddy},
\begin{eqnarray}
\nonumber
\lambda^2-2(a+i\omega)\lambda+(a^2-\omega^2+i 2 a \omega)-
\ep^2 e^{-2\lambda\tau}=0,\\
\label{eq:eigen}
\end{eqnarray}
where $a=1-\ep$.
                                                                     
Letting $\lambda=\alpha+i\beta$ where $\alpha$ and $\beta$ are
real and imaginary part of the eigenvalues
in Eq.~(\ref{eq:eigen}), and separating real and imaginary parts  
leads to the equations
\begin{eqnarray}
\nonumber
\alpha^2-\beta^2-2(a\alpha-\beta\omega)+a^2-\omega^2-\ep^2e^{
-2\alpha\tau} \cos{(2\beta\tau)}&=&0\\
&&
\label{eq:freq1}
\end{eqnarray}
and
\begin{eqnarray}
\nonumber
\label{eq:freq2}
2 \alpha\beta-2(\alpha\omega+a\beta)+2a\omega+\ep^2e^{
-2\alpha\tau} \sin{(2\beta\tau)}&=&0.\\
&&
\end{eqnarray}

As these equations are  difficult to solve analytically
we use the Lyapunov exponent of the fixed point, $\lambda_{FP}$, 
to find $\beta$ numerically.  The Lyapunov exponent of a fixed point is
equal to the real part of the eigenvalue, \ie $\lambda_{FP}=\alpha$, 
we use this in Eqs.~(\ref{eq:freq1}) and (\ref{eq:freq2}). The 
resulting roots of these equations give $\beta$ and 
are shown separately by circles and stars in Fig.~\ref{fig:reddy}(g).
The final solution which satisfying both these equations will be 
the solution of the eigenvalue equation, Eq.~(\ref{eq:eigen}).
Here we see that prior to the transition, the smaller root
of Eq.~(\ref{eq:freq1}) matches Eq.~(\ref{eq:freq2}) while 
after the transition, it is the larger root. 
The agreement with $\beta$ computed directly from the eigenvalue 
equation and numerically calculated frequency (dashed line)  is excellent.  

This analysis has been possible for the case of coupled limit cycle
oscillators. Although it cannot be extended to the case of chaotic
systems, it is possible to conjecture that the underlying mechanism
could be similar, namely that the transition is due to a jump in 
the imaginary part of the eigenvalue of the fixed point.

\section{Discussion and Summary}

In this paper I have shown that the phenomenology developed for the 
cause of amplitude death in systems with limit--cycle 
oscillators \cite{eli,am,stro} can  be extended to the case of chaotic 
dynamical systems.  This also extends earlier work on amplitude 
modulation  \cite{pla}, where time-delay coupling
has been used to stabilize undesirable low-frequency chaotic 
fluctuations in a semiconductor laser. (There it was seen there that time-delay 
coupling could convert  chaotic oscillations of the laser field with power
dropouts into quasiperiodic motion without power dropouts.)

Amplitude death in chaotic systems is a consequence of 
time--delays that stabilize fixed points that might either have 
existed in the uncoupled system, or that might have been created 
through the interaction. The  evidence  presented here is largely numerical.
We considered separately  the cases of 
the interacting systems being (a) identical, (b) the same but with 
mismatched parameters, and (c) completely distinct. In all cases, 
amplitude death can occur, and moreover, this can occur over a range 
of parameter values.  The mechanics through which chaos converts into
fixed point motion are different in these example: some cases it happens
via limit-cycle at Hopf bifurcation while other cases 
fixed point gets stablised directly. In mixed systems
this stabilization is initiated via riddling phenomenon.
 
We have used the  \le as indicator to detect 
the different dynamical regimes, and to detect transitions, as 
for example, in the case of  identical oscillators, when the 
subsystems are  oscillating in or out of phase with each other.
When the subsystems are maximally stable, namely the largest \le is minimum,
a ``phase'' transition occurs, and motion goes from being in phase to being
out of phase.  Measures that detect this transition are the phase difference 
and frequency; although not possible to analyze in chaotic systems, it
is possible to get some understanding of this transition for a limit--cycle 
case. Apart from its mathematical interest, the possibility of 
using a transition from dynamics on a limit cycle which is in 
phase to that which is out of phase in practical applications 
suggest itself. 
In experimental application of  amplitude death phenomena,
where transient dynamics always persists,
 it will be important to know the  parameters
where motion is either in-phase (low frequency) or out of
phase (high frequency). An important possible use of it
could be a  coupled chaotic lasers \cite{pla},
where relatively higher degree of constant output could be
 obtained by keeping the parameters of
the lasers and the time-delay coupling in the out of phase 
  regime compared to that of the  in-phase \cite{new}.

Experimental verification of this generalization of the conventional 
amplitude-death phenomenon that occurs in limit-cycle oscillators may 
in fact, prove to be technically simple to achieve since it persists 
even after  parameters mismatch, and occurs for different types of coupled 
oscillators, bypassing the need for  irregular oscillations 
(\eg Ref.~\cite{pla}).  

\begin{acknowledgments} I thank Ram Ramaswamy for discussions and comments,
and anonymous referees of an earlier version of the manuscript for very
helpful criticism. A part of this work was supported  by the Department of Science and Technology, India.
\end{acknowledgments}

\eject

\begin{thebibliography}{99}
\bibitem{num} Fig.~\ref{fig:ident} is obtained by computing the largest
Lyapunov exponent for the system and examining regions where it is 
positive or negative in a $100\times 100$ grid.
For calculating the Lyapunov exponents \cite{farmer} the same initial 
conditions are taken at each point in the phase diagram. The flow is 
integrated using the  Runge-Kutta 4th order scheme  with integration 
step $\Delta t=\tau/N$ where  $N=300$ is fixed (making the 
dimensionality of the system 6$N$ due to the delay).
\bibitem{traj} 
In order to show the effect of the interaction, we project the orbit
of the system onto the coordinates of the individual oscillators 
separately.
\bibitem{trans} Measurement of the phase difference or
the common frequency is done when amplitude death has set
in and the orbit starts spiraling towards the fixed point.
\bibitem{phasediff} Defining a phase in  chaotic oscillators can
be difficult, but as has been suggested \cite{rosen}, by examining
projections in planes $x_1-x_2$ or $y_1-y_2$,
we can approximate phase as $\theta_1 \sim \tan^{-1}(x_2/x_1)$ or
 $\theta_2 \sim \tan^{-1}(y_2/y_1)$.
\bibitem{lor} In this system, Eq.~(\ref{eq:lorenz}), there are two
fixed points which are stable in FP region, \ie there are coexisting
attractors. The stability of one of these is shown in  
Fig.~\ref{fig:loren}(b). In ongoing work, a detailed exploration of
the boundaries of these attractor basins in the presence of time delay 
is being studied.
\bibitem{phase} We use the term `phase' in the region of
\am even though the amplitude of oscillation decreases with 
time \eg the inset Figs.~\ref{fig:diff}(e) and (f) which
clearly indicate that the trajectories are in or
out of phase.
\end{thebibliography}
\end{document}